\documentclass{emulateapj}
\shorttitle{Analysis of Weibel filaments}
\shortauthors{SUZUKI \& SHIGEYAMA}
\begin{document}
\title{DETAILED ANALYSIS OF FILAMENTARY STRUCTURE IN THE WEIBEL INSTABILITY}
\author{AKIHIRO SUZUKI\altaffilmark{1,2} and TOSHIKAZU SHIGEYAMA\altaffilmark{1}}
\altaffiltext{1}{Research Center for the Early Universe, School of Science, University of Tokyo, Bunkyo-ku, Tokyo 113-0033, Japan.}
\altaffiltext{2}{Department of Astronomy, School of Science, University of Tokyo, Bunkyo-ku, Tokyo 113-0033, Japan.}
\begin{abstract}
We present results of a 2D3V kinetic Vlasov simulation of the Weibel instability.
The kinetic Vlasov simulation allows us to investigate the velocity distribution of dilute plasmas, in which the effect of collisions between particles is negligible, and has the advantage that the accuracy of the calculated velocity distribution does not depend on the density of plasmas at each point in the physical space.  
We succeed in reproducing some features of the Weibel instability shown by other simulations, for example, the exponentially growing phase, the saturation of the magnetic field strength, the formation of  filamentary structure, and the coalescence of the filaments. 
Especially, we concentrate on the behavior of the filaments after the saturation of the magnetic field strength and find that there is a kind of quasi-equilibrium states before the coalescence occurs. 
Furthermore, it is found that an analytical solution for stationary states of the 2D3V Vlasov-Maxwell system can reproduce some dominant features of the quasi-equilibrium, e.g, the configuration of the magnetic field and the velocity distribution at each point. 
The analytical expression could give a plausible model for the transition layer of a collisionless shock where a strong magnetic field generated by the Weibel instability provides an effective dissipation process instead of collisions between particles. 
\end{abstract}
\keywords{plasmas --- instabilities --- shock waves --- magnetic fields}

\section{INTRODUCTION}
It has been a long time since the existence of collisionless shocks was confirmed. 
In this context, the collisionless shock means discontinuities of physical quantities occurring in sufficiently rarefied plasmas, in which the effect of collisions between particles composing the plasma is negligible. 
So far, it is believed that collisionless shocks play important roles in a wide class of astrophysical phenomena, e.g., supernova remnants(SNRs), the prompt emission and afterglow of gamma-ray bursts(GRB), jets from active galactic nuclei. 

From the discovery of the bow shock in front of the terrestrial magnetosphere  \citep{n64}, the properties of collisionless shocks, e.g., the mechanism of the formation, the shock conditions, and the configuration of the magnetic field, have been investigated by various approaches. 
For example, \cite{p65} observed the formation of a collisionless shock in laboratory experiments. 
Observationally, several satellites have revealed the nature of collisionless shocks. 
Especially, GEOTAIL measured the variation of the magnetic field strength in the bow shock in front of the terrestrial magnetosphere from the upstream to the downstream. GEOTAIL also acquired the energy spectra of ions in the solar wind, which confirmed the existence of non-thermal particles.  

Also in other high-energy astrophysical phenomena such as SNRs and GRBs, the existence of non-thermal particles is confirmed observationally. 
In generating such particles, a strong magnetic field amplified around the collisionless shock is considered to be a key ingredient. 
This prospect is supported by observations of several SNRs\citep{vl03,b03}. 
Additionally, the recent observation of the time variation of X-ray sources in RX J1713.7-3946 by \cite{u07} strongly indicates the existence of a strong magnetic field in the shock front. 
Also in GRB afterglows, the existence of a long-lived, near equipartition magnetic field is indispensable for the observed synchrotron emissions\citep[see, e.g,][]{p99,mes06}. 

Theoretically, a great deal of investigation has been done in both analytical and numerical framework. 
\cite{ml99} suggested that the Weibel instability\citep[][]{w59,f59}, which is a kind of plasma instability caused by the anisotropic velocity distributions of collisionless plasmas, could dissipate the kinetic energy of the plasma and generate a strong magnetic field in the transition layer of a collisionless shock. 
First principle simulations of the formation of a collisionless shock \citep[][]{k07,sp08,sp08b} by using recent powerful computers have supported the suggestion. 
Especially, \cite{sp08b} performed long-term 2D particle-in-cell(PIC) simulations of a relativistic collisionless shock in an unmagnetized pair plasma and found a sign of the generation of non-thermal particles via the first-order Fermi acceleration mechanism\citep[][]{f49}. 
They conclude that the scattering source of particles, which is needed for the Fermi acceleration to occur, is a turbulent magnetic field generated by the Weibel instability in the transition layer of the shock. 
This is why the Weibel instability is a compelling candidate for the mechanism of the formation of the shock front associated with a strong magnetic field and particle acceleration. 

However, there remain some problems regarding the Weibel instability itself. 
Results of recent 3D simulations of the Weibel instability \citep[][]{s03,f04} show that filamentary structure (sometimes called "current filament") forms as a magnetic field is amplified. 
After saturation of the magnetic field strength, the filaments coalesce each other to form larger one. 
Although each filament is small (its initial radius is roughly equal to a few electron skin depth), the detailed properties of this structure, e.g., the correlation length of the magnetic field, the particle energy distribution, and the time evolution, may affect the macroscopic behaviors of the transition layer of a collisionless shock.
For example, \cite{m00} pointed out a possibility that ultrarelativistic electrons accelerated by highly nonuniform small scale magnetic fields emit radiations different from a synchrotron radiation emitted by electrons in uniform magnetic fields. 
Subsequently some investigations into the filamentary structure have been performed.
From the conservation of energy, \cite{g01} predicted that the magnetic field energy decreases as $t^{-1}$. 
\cite{csa08} considered a variant of Landau damping (a process dissipating the energy of electromagnetic fields into the kinetic energy of particles) and reached a similar result for the evolution of the magnetic fields. 
They assumed straight orbits of particles because of weak magnetic fields, though their PIC simulations reveal that there exist some regions where strong magnetic fields bend the orbits of particles. 
\cite{m05} investigated the coalescence of the filaments by using  a two-dimensional toy model.
They pointed out that the correlation length of the magnetic field grows exponentially at first and then linearly with time. 
\cite{awn07} reexamine the toy model including the effect of screening currents of the background plasma and predict that the coalescence slows down when the separation of the filaments becomes larger than the skin depth of the background plasma. 
But their analysis does not predict whether there exists a critical scale length above which the coalescence does not proceed. 
Furthermore, their simple model ignores micro processes such as the dissipation of the magnetic energy into the kinetic energy by the reconnection of the magnetic field. 
On the other hand, \cite{m06} treated the filamentary structure in the framework of the magnetohydrodynamics(MHD). 
They argue that a pressure-driven MHD instability destroys the filament and the strong magnetic field generated in the transition layer is short-lived. 
That is, the consensus is yet to be reached regarding the long-term evolution of the Weibel filaments. 
Especially, it is unclear whether the toy model and the MHD model are appropriate to model the filaments. 

To settle such unresolved problems, we use a kinetic Vlasov code. 
This approach enables us to investigate the structure of the filaments and their coalescence and has the advantage that the accuracy of the calculated velocity distribution does not depend on the density of plasmas at each point in the physical space, while it does in PIC simulations. 
However, there are some computational constrains on Vlasov codes. 
The most serious one is its small dynamic range in both physical and velocity spaces. 
Due to the defect, we cannot treat the formation of a shock from two plasma flows or relativistic motions of particles. 
Thus, in this work, we concentrate on the coalescence of filaments forming as a result of the non-relativistic Weibel instability. 
The saturation of the relativistic Weibel instability is realized by the same mechanism as that of the non-relativistic Weibel instability \citep{kt08}. 
Therefore, if we can construct a reliable model of the filaments by analyzing the non-linear behavior of the non-relativistic Weibel instability in detail, it may provide a general insight into some essential processes operating in the transition layer of a collisionless shock and predict their long-term evolution beyond the reach of the simulation. 

This paper is organized as follows. 
In the next section, we describe the strategy for the numerical simulation. 
The results are shown in $\S$3. 
In $\S$4 , we construct an analytic model for the filaments and address remaining problems. 
We conclude this paper in $\S$5.

\section{FORMULATION}
In this section, we describe the governing equations, some assumptions, and an initial setup for the numerical simulation. 
\subsection{Equations}
The equations describing the behavior of rarefied plasmas are the Vlasov-Maxwell system \citep[see, e.g.,][]{s94}. 
The Vlasov equation governs the time evolution of the distribution function $f_j(t,x,y,z,v_x,v_y,v_z)$ of species $j$($i$ for ions and $e$ for electrons). Here the cartesian coordinates are $(x,y,z)$ and the corresponding coordinates in the velocity space are $(v_x,v_y,v_z)$.  
The Maxwell equations govern the time evolution of the electromagnetic fields ${\bf E}$ and ${\bf B}$. 
We impose two assumptions. 
One is that the plasma is homogeneous in the $z$ direction, which makes the Vlasov equation to take the following form;
\begin{equation}
\frac{\partial f_j}{\partial t}+v_x\frac{\partial f_j}{\partial x}+v_y\frac{\partial f_j}{\partial y}+
\frac{q_j}{m_j}\left({\bf E}+\frac{\bf v}{c}\times{\bf B}\right)\cdot\frac{\partial f_j}{\partial {\bf v}}=0,
\label{vlasov}
\end{equation}
where $q_j$ and $m_j$ are the charge and the mass of species $j$, and $c$ is the speed of light. 
${\bf E}$ and ${\bf B}$ are expressed by introducing the scalar and the vector potentials, $\phi(t,x,y)$ and ${\bf A}(t,x,y)$ as 
\begin{equation}
{\bf E}=-\nabla \phi -\frac{1}{c}\frac{\partial{\bf A}}{\partial t},\ \ \ 
{\bf B}=\nabla\times{\bf A}.
\end{equation}
The time evolution of the potentials is described by the wave equations;
\begin{equation}
\frac{1}{c^2}\frac{\partial^2\phi}{\partial t^2}-\nabla^2\phi=-4\pi\rho,\ \ \ 
\frac{1}{c^2}\frac{\partial^2{\bf A}}{\partial t^2}-\nabla^2{\bf A}=-4\pi{\bf j},
\label{wave}
\end{equation} 
where $\rho$ and ${\bf j}$ are the electric density and the electric current density, which satisfy the Lorenz condition;
\begin{equation}
\frac{1}{c}\frac{\partial \phi}{\partial t}+\nabla\cdot{\bf A}=0.
\label{Lorenz}
\end{equation}
The other assumption is that the plasma consists of ions and electrons with the same charge but with the opposite sign ($q_i=-q_e=e$). 
So the source terms in Equation (\ref{wave}) are expressed in terms of $f_j$ as 
\begin{eqnarray}
\rho&=&e\int^\infty_{-\infty}\int^\infty_{-\infty}\int^\infty_{-\infty}
(f_i-f_e)dv_xdv_ydv_z,\\
{\bf j}&=&\frac{e}{c}\int^\infty_{-\infty}\int^\infty_{-\infty}\int^\infty_{-\infty}{\bf v}
(f_i-f_e)dv_xdv_ydv_z.
\label{source}
\end{eqnarray}

\subsection{Normalization}
We define some values characterizing the physical quantities; $1/\omega_\mathrm{e}$ as the time scale, $c/\omega_\mathrm{e}$ as the length scale, $c$ as the velocity, and $m_\mathrm{e}c\omega_\mathrm{e}/e$ as the electromagnetic field. 
Here $\omega_e$ is the electron plasma frequency. 
Normalized by these values, the governing equations described in the previous subsection can be expressed in a non-dimensional form. 
In the following, we treat thus normalized physical variables. 

\subsection{Method of numerical integration}
To simulate the Weibel instability, we use the numerical scheme for the integration of the Vlasov-Maxwell system proposed by \cite{mcct02}. 
We review their procedure that evolves the distribution functions and the electromagnetic potentials by a time interval $\Delta t$ in the following steps. 

In integrating the Vlasov equation (\ref{vlasov}), we treat the electromagnetic field as static. 
One can see the Vlasov equation (\ref{vlasov}) consists of five one-dimensional advection equations in the form of
\begin{equation}
\frac{\partial g}{\partial t}+a_\zeta\frac{\partial g}{\partial \zeta}=0,
\label{adv}
\end{equation}
where $\zeta(=x,y,v_x,v_y,v_z)$ is an independent variable, $a_\zeta$ is a constant and $g$ is 
a function of $(t, \zeta)$. 
This equation is solved by second-order van Leer's scheme. 
Here we define an operator ${\cal T}$ that evolves the function $g$ by a time interval $\Delta t$ according to Equation (\ref{adv});
\begin{equation}
g(t+\Delta t,\zeta-a_\zeta\Delta t)={\cal T}[\zeta,\Delta t]g(t,\zeta).
\end{equation}
Using this operator, \cite{mcct02} proposed the following scheme to integrate the Vlasov equation (\ref{vlasov});
\begin{eqnarray}
f_j(t+\Delta t,{\bf x},{\bf v})&=&{\cal S}[x,y,\Delta t/2]{\cal S}[v_x,v_y,\Delta t/2]
\nonumber\\
&&\times{\cal T}[z,\Delta t]{\cal S}[v_x,v_y,\Delta t/2]
\nonumber\\
&&\times{\cal S}[x,y,\Delta t/2]f_j(t,{\bf x},{\bf v}).
\end{eqnarray}
where another operator ${\cal S}$ has been defined as
\begin{equation}
{\cal S}[x,y,\Delta t]\equiv {\cal T}[x,\Delta t/2]{\cal T}[y,\Delta t]{\cal T}[x,\Delta t/2].
\end{equation}

If the distribution function at $t+\Delta t$ is given, the source terms in Equations (\ref{wave}) are calculated by Equation (\ref{source}). 
The wave equations (\ref{wave}) governing the time evolution of the electromagnetic fields are solved by using the Fourier transformation, which imposes the periodic boundary conditions in the $x$ and $y$ directions implicitly;
\begin{eqnarray}
\hat{h}(t,k_x,k_y)&=&\int^{L_x/2}_{-L_x/2}\int^{L_y/2}_{-L_y/2}\exp[-i(k_xx+k_yy)]
\nonumber\\
&&\times h(t,x,y)dxdy, 
\label{fourier}
\end{eqnarray}
where $L_x(L_y)$ is the length of the computational domain in the $x(y)$ direction and $h$ is a function of $(t,x,y)$. The inverse Fourier transformation is defined by
\begin{eqnarray}
h(t,x,y)&=&\frac{1}{(2\pi)^2L_xL_y}\int^{L_x/2}_{-L_x/2}\int^{L_y/2}_{-L_y/2}\exp[i(k_xx+k_yy)]
\nonumber\\
&&\times\hat{h}(t,k_x,k_y)dxdy.
\label{inverse}
 \end{eqnarray} 
 These transformations are computed by the standard Fast-Fourier-Transformation scheme \citep[see, e.g.][]{recipe} numerically.
Substitution of these relations for $\phi,{\bf A},\rho,{\bf j}$ into Equation (\ref{wave}) leads to the following four second-order ordinary differential equations;
\begin{equation}
\frac{d^2\hat{\phi}}{dt^2}=-(k_x^2+k_y^2)\hat{\phi}-4\pi\hat{\rho},\ \ \ 
\frac{d^2\hat{\bf A}}{dt^2}=-(k_x^2+k_y^2)\hat{\bf A}-4\pi\hat{\bf j},
\label{eq12}
\end{equation}
which are solved by the Runge-Kutta method of order 4. 
To ensure that the obtained $\hat{\phi}(t)$ and $\hat{\bf A}(t)$ satisfy the Lorenz condition (\ref{Lorenz}), we use the values of $d\hat{\phi}/dt$ calculated from (\ref{Lorenz}) with ${\bf A}(t)$ instead of using those obtained by numerically integrating equations (\ref{eq12}).

In that way, we evolve the velocity distributions and the electromagnetic field one after the other.

\subsection{Simulation setups}
As the initial condition, we consider a counter-streaming plasma that is homogeneous in space;
\begin{eqnarray}
f_{j0}&=&\frac{n_j}{2\pi^{3/2}v_{j}^3}\left\{\exp\left[-\frac{v_x^2+v_y^2+(v_z+V_j)^2}{v_j^2}\right]\right.
\nonumber\\
&&\left.+\exp\left[-\frac{v_x^2+v_y^2+(v_z-V_j)^2}{v_j^2}\right]\right\}, 
\end{eqnarray}
where $n_j$, $v_j$, and $V_j$ are the number density, the thermal velocity, and the bulk velocity of species $j$. We assume that they are constants and $n_\mathrm{i}=n_\mathrm{e}=n_0$ for the charge neutrality. 
In our simulation, the values of these parameters are as follows; 
$n_0=1$, $v_\mathrm{e}=0.05$, $v_\mathrm{i}=0.05\sqrt{m_\mathrm{e}/m_\mathrm{i}}$, and $V_\mathrm{e}=V_\mathrm{i}=0.2$. 
The mass ratio is assumed to be $m_\mathrm{i}/m_\mathrm{e}=1$ and $16$. 
The initial configuration of the electromagnetic field is
\begin{equation}
\phi=A_x=A_y=0,\ \ \ A_z=-\epsilon\sin x\sin y,
\label{perturbation}
\end{equation}
which is equivalent to
\begin{eqnarray}
&&E_x=E_y=E_z=0,
\nonumber\\
&&B_x=\epsilon \sin x\cos y,\ \ \ B_y=-\epsilon\cos x\sin y,\ \ \ B_z=0,
\end{eqnarray}
where we have introduced a small parameter $\epsilon(=10^{-5})$. 
In other words, we treat the above magnetic field as a perturbation to the initial distribution of particles with no electromagnetic field. 
Next, we address the boundary conditions. 
The simulation domain consists of the spatial intervals given by $x,y\in[-\pi,\pi]$ and the velocity intervals given by  $v_x,v_y\in[-0.4,0.4]$ and $v_z\in [-0.6,0.6]$.
The periodic boundary condition is imposed in the $x$ and $y$ direction, while in the velocity space, the distribution function assumed to vanish for $|v_x|, |v_y|>0.4$ and $|v_z|>0.6$. 
The number of zones in the physical space is $32\times 32$ and that in the velocity space is $40\times 40\times 60$. 

\section{RESULTS}
In this section, we show the results of the integration of Equations (\ref{vlasov})-(\ref{source}) under the initial condition given in the previous section. 

\subsection{The time evolution of energies}
\begin{figure}
\begin{center}
\includegraphics[scale=0.5]{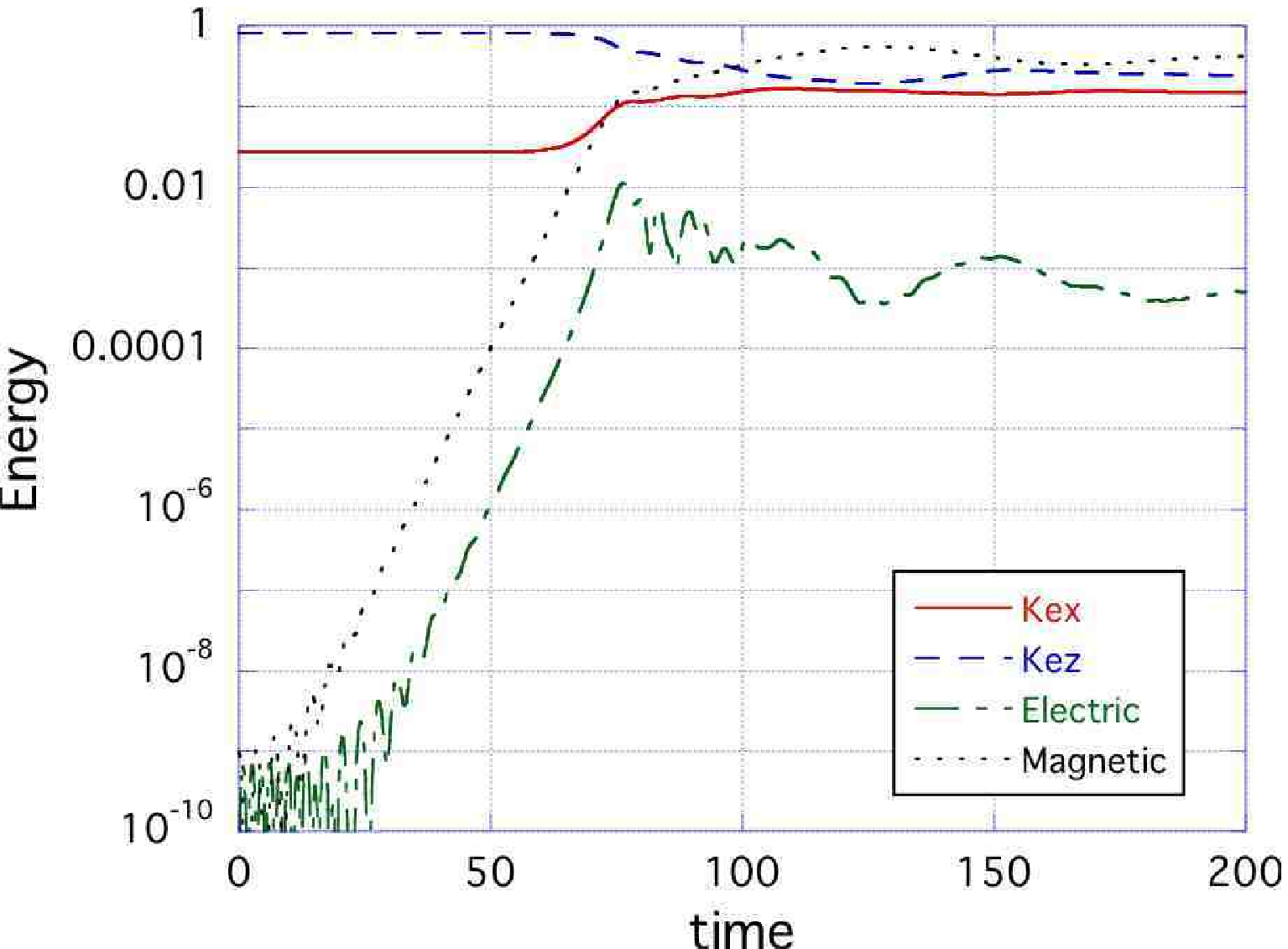}
\caption{The time evolution of the electron kinetic energies, $K_{ex}$ (solid), $K_{ez}$ (dashed), the electric energy $E_\mathrm{e}$ (dash-dotted), and the magnetic energy $E_\mathrm{m}$ (dotted), for $0<t<200$ and $m_i/m_e=16$.}
\label{figure1}
\end{center}
\end{figure}

Figure \ref{figure1} shows the time evolution ($0<t<200$ for the case $m_\mathrm{i}/m_\mathrm{e}=16$) of the electron kinetic energies in the $x$ and $z$ directions, the electric energy, and the magnetic energy. 
The kinetic energy of species $j$ in each direction is defined by
\begin{eqnarray}
K_{jx}&=&\frac{m_j}{2}\int^\pi_{-\pi}\int^\pi_{-\pi}\int^\infty_{-\infty}\int^\infty_{-\infty}\int^\infty_{-\infty}v_x^2f_j
\nonumber\\
&&\hspace{10em}\times dxdydv_xdv_ydv_z,\\
K_{jy}&=&\frac{m_j}{2}\int^\pi_{-\pi}\int^\pi_{-\pi}\int^\infty_{-\infty}\int^\infty_{-\infty}\int^\infty_{-\infty}v_y^2f_j
\nonumber\\
&&\hspace{10em}\times dxdydv_xdv_ydv_z,\\
K_{jz}&=&\frac{m_j}{2}\int^\pi_{-\pi}\int^\pi_{-\pi}\int^\infty_{-\infty}\int^\infty_{-\infty}\int^\infty_{-\infty}v_z^2f_j
\nonumber\\
&&\hspace{10em}\times dxdydv_xdv_ydv_z,
\end{eqnarray}
while the electric and the magnetic energies are defined by
\begin{eqnarray}
E_\mathrm{e}&=&\frac{1}{2}\int^\pi_{-\pi}\int^\pi_{-\pi}(E_x^2+E_y^2+E_z^2)dxdy,\\
E_\mathrm{m}&=&\frac{1}{2}\int^\pi_{-\pi}\int^\pi_{-\pi}(B_x^2+B_y^2+B_z^2)dxdy.
\label{Em}
\end{eqnarray}
\begin{figure}
\begin{center}
\includegraphics[scale=0.5]{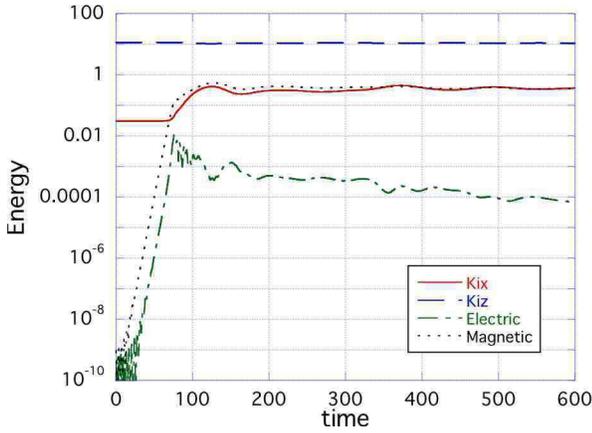}
\caption{The time evolution of the ion kinetic energies, $K_{ix}$ (solid), $K_{iz}$ (dashed), the electric energy $E_\mathrm{e}$ (dash-dotted), and the magnetic energy $E_\mathrm{m}$ (dotted), for $0<t<600$ and $m_i/m_e=16$.}
\label{figure2}
\end{center}
\end{figure}

The values of $K_{jx}$ and $K_{jy}$ evolve in the same way, because of the symmetry in the $v_x$-$v_y$ plane. 
The time evolution is divided into two phases, the linear phase and the saturation phase. 
In the linear phase, the electric and magnetic energies increase exponentially. 
The growth rate is consistent with a linearized analysis \citep{cpbm98}. 
After the linear phase, non-linear effects become significant and the magnetic energy becomes comparable to the total kinetic energy of electrons, while the electric energy takes relatively small values. 
In the transition between the two phases, $K_{ex}$ increases and $K_{ez}$ decreases. 
This is one of the remarkable features of the Weibel instability, the anisotropy of the motions is mediated by interactions between particles and electromagnetic fields. 

From the time evolution of the ion kinetic energy in each direction and the electric and magnetic energies shown in Figure \ref{figure2}, it is found that its large mass ($m_\mathrm{i}=16m_\mathrm{e}$) keeps the value of $K_{iz}$ almost constant by inhibiting the efficient exchange between the kinetic energy and the electromagnetic energy. 

\begin{figure}
\begin{center}
\includegraphics[scale=0.5]{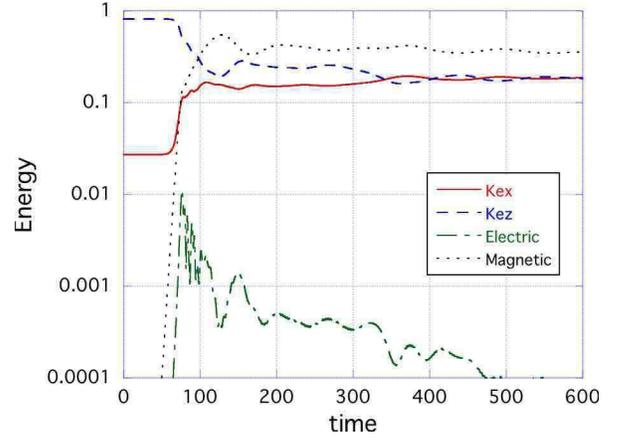}
\caption{Same as Figure \ref{figure1}, but for $0<t<600$.}
\label{figure3}
\end{center}
\end{figure}

Figure \ref{figure3} shows the time evolution ($0<t<600$) of $K_{ex}$, $K_{ez}$, $E_\mathrm{e}$, and $E_\mathrm{m}$. 
In the saturation phase, the electric energy gradually decreases indicating that the difference between the density distributions of ions and electrons diminishes. 
However, it does not mean that the contribution of the electric force to the force balance is negligible, 
because the balance of the Lorentz force acting on a charged particle is governed by the following relation;
\begin{equation}
{\bf E}+\langle{\bf v}_j\rangle\times{\bf B}=0,
\end{equation}
where $\langle{\bf v}_j\rangle$ is the bulk velocity of species $j$ defined by
\begin{equation}
\langle{\bf v}_j\rangle=\int^\infty_{-\infty}\int^\infty_{-\infty}\int^\infty_{-\infty}{\bf v}f_jdv_xdv_ydv_z.
\end{equation}
In this simulation, a small value of the bulk velocity of ions of the order of $|\langle{\bf v}_\mathrm{i}\rangle|\sim 0.1$ enables the electric field to contribute to the achievement of the force balance. 

\subsection{The saturated value of the magnetic field strength}
\begin{figure*}
\includegraphics[scale=0.9]{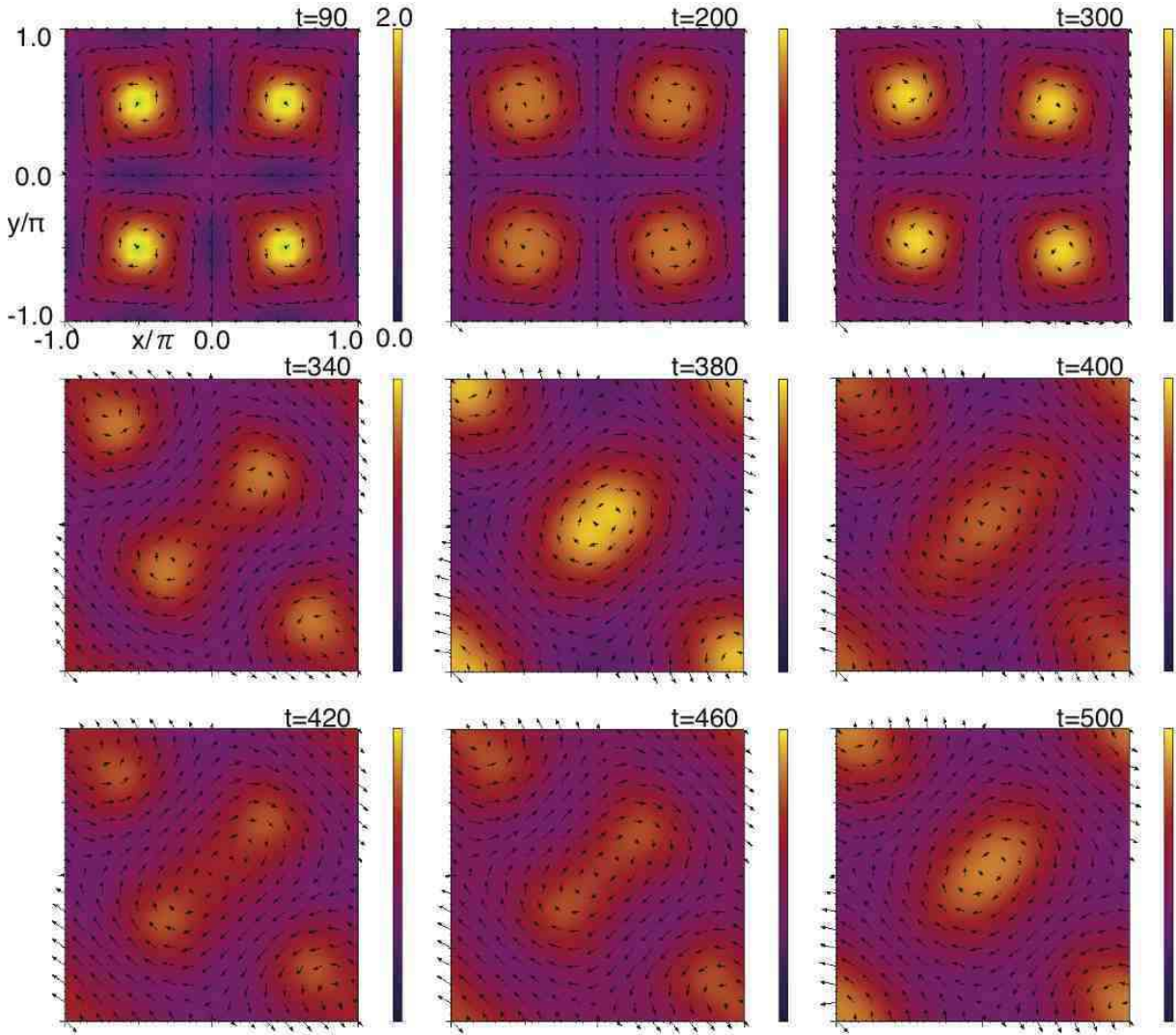}
\caption{Snapshots of the time evolution of the Weibel instability. Each panel represents a color-coded density distribution of electrons(or ions) and the magnetic fields by arrows in the x-y plane.}
\label{figure4}
\end{figure*}
Several authors \citep[][]{cpbm98,ml99,k05} have investigated the saturation mechanism of the Weibel instability. 
Their scenario is as follows. 
If the orbits of particles are straight, the particles can form current filaments, which generate magnetic fields around themselves. 
Thus the magnetic field strength grows consistently as long as the orbits of particles can be regarded as straight one. 
On the other hand, once the magnetic field strength becomes strong enough to deflect the orbits of particles, the particles cannot form current filaments. 
Hence, it is expected that the magnetic field strength will be saturated when the Larmor radius $r_\mathrm{L}$ of a particle becomes comparable to the characteristic scale length $l$ of the plasma. 
\cite{cpbm98} adopted the electron skin depth as $l$, while \cite{ml99} adopted the reciprocal of the wave number of the most unstable mode of the Weibel instability as $l$. 
In \cite{k05}, results of a series of PIC simulations imply that the radius of a filament is compatible to the characteristic scale length.  
In our simulation, the wave number of the growing mode is fixed {\it ab initio} from the form of the perturbation (\ref{perturbation}) and identical to the electron skin depth $c/\omega_\mathrm{e}$, which leads to the formation of filaments with radius $R=\pi/2$. 
Then, we obtain the following condition for the saturation of the magnetic field strength;
\begin{equation}
r_\mathrm{L}=\frac{V_\mathrm{e}}{B_\mathrm{sat}}=\frac{\pi}{2}.
\end{equation} 
We can estimate the saturated value of the magnetic field strength $B_\mathrm{sat}=2v_\mathrm{e}/\pi =0.13$ by substituting $V_\mathrm{e}=0.2$. 
The corresponding magnetic field energy is $E_\mathrm{m}=B_\mathrm{sat}^2L_xL_y/2= 0.32$, while $E_\mathrm{m}\simeq 0.4$ for $m_\mathrm{i}=16m_\mathrm{e}$ and $E_\mathrm{m}\simeq0.2$ for $m_\mathrm{i}=m_\mathrm{e}$ from the results of our simulation. 
Therefore, this rough estimation of the saturated value of the magnetic field strength is in good agreement with the results of the simulation.

\subsection{The density distributions}

Figure \ref{figure4} shows snapshots of the electron (or ion) density distribution and the magnetic field configuration at $t=90, 200,300,340,380,400,420,460,500$. 
These trace the time evolution of the filamentary structure resulting from the development of the Weibel instability. 
In the linear phase, the filamentary structure develops according to the initial configuration of the magnetic field. 
In the beginning of the saturation phase ($90<t<300$), we can see two filaments carrying electric currents parallel to the $z$-axis around $(x,y)=(-\pi/2,-\pi/2)$, $(\pi/2,\pi/2)$ and the two other filaments carrying anti-parallel currents around $(x,y)=(\pi/2,-\pi/2)$, $(-\pi/2,\pi/2)$. 
In this period, a quasi-equilibrium configuration seems to be achieved, since the distribution function hardly changes seems to be achieved. 
After a while, the configuration changes in the following way. 
Adjacent two filaments carrying currents to the same direction begin to approach each other ($300<t<340$) and coalesce into a large filament ($340<t<350$). 
As the snapshots at $t=400,420,460,500$ exhibit, the large filament oscillates after the coalescence. 
The amplitude of the oscillation gradually decreases to achieve a new equilibrium configuration, which is self-similar to the previous one. 

\subsection{Thermalization of electrons}
\begin{figure}
\begin{center}
\includegraphics[scale=0.5]{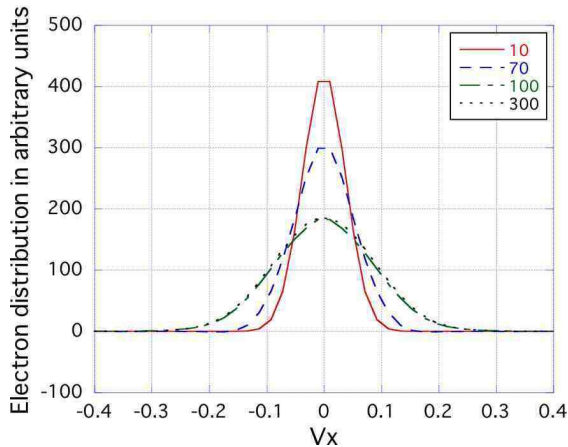}
\caption{The $x,y,v_y,v_z$-integrated electron velocity distribution at $t=10,70,100$, and $300$.}
\label{figure5}
\end{center}
\end{figure}

As we see in the previous subsections, the Weibel instability amplifies a transverse electromagnetic field by rapidly dissipating the bulk kinetic energy of particles. 
However, all the dissipated energy is not converted into the electromagnetic energy.  
It is known that the electromagnetic field thus amplified efficiently thermalize electrons. 
We can see this effect in Figure \ref{figure5}, where each line represents the $x,y,v_y,v_z$-integrated velocity distribution of electrons at $t=10, 70, 100,$ and $300$. 
Because the velocity distributions in Figure \ref{figure5} are well represented by the Maxwell-Boltzmann distribution with zero bulk velocity, the kinetic energies $K_{\mathrm{e}x}$ in Figures \ref{figure1} and \ref{figure3} are dominated by that of the thermal velocity. 

\begin{figure}
\begin{center}
\includegraphics[scale=0.5]{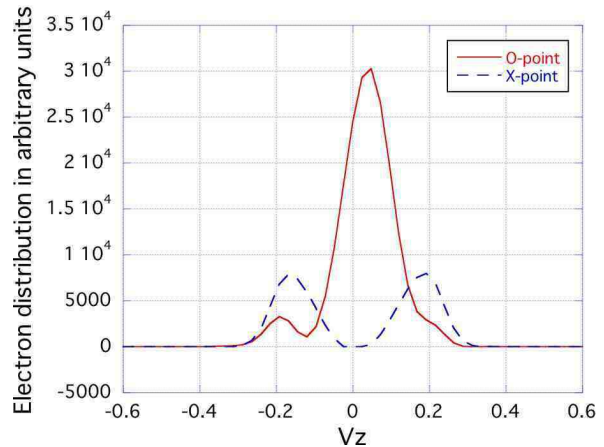}
\caption{The electron velocity distribution at X-point(solid) and O-point(dashed) in arbitrary units for $m_\mathrm{i}/m_\mathrm{e}=16$ and $t=280$.}
\label{figure6}
\end{center}
\end{figure}

\subsection{The velocity distributions in the quasi-equilibrium}
\begin{figure}
\begin{center}
\includegraphics[scale=0.5]{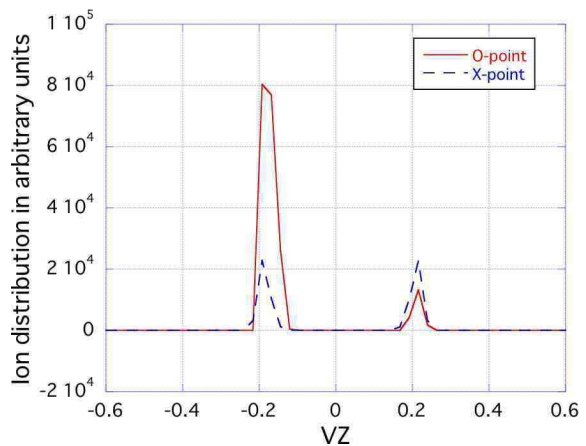}
\caption{The ion velocity distribution at X-point(solid) and O-point(dashed) in arbitrary units for $m_\mathrm{i}/m_\mathrm{e}=16$ and $t=280$.}
\label{figure7}
\end{center}
\end{figure}

\begin{figure}
\begin{center}
\includegraphics[scale=0.5]{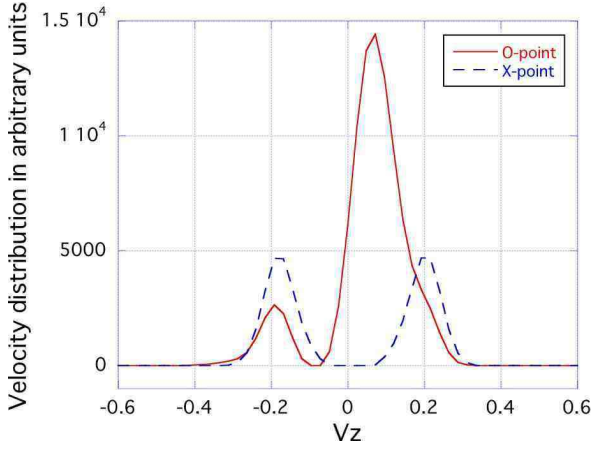}
\caption{The electron(or ion) velocity distribution in the quasi-equilibrium at X-point(solid) and O-point(dashed) in arbitrary units for $m_\mathrm{i}/m_\mathrm{e}=1$. From the mass symmetry, the ion velocity distribution at any point become identical with that of electron at the same point by transforming $v_z$ as $v_z\ \rightarrow\ -v_z$}
\label{figure8}
\end{center}
\end{figure}

In the first quasi-equilibrium ($200<t<300$), we focus on the velocity distribution at two points. 
They are the point at the center of a filament located at $(x,y)=(\pi/2,\pi/2)$ (referred to as O-point), and  the point surrounded by four filaments at $(x,y)=(0,0)$ (referred to as X-point). 
In Figure \ref{figure6}, solid and dashed lines represent the $v_x,v_y$-integrated electron velocity distributions in the period of the quasi-equilibrium at the O- and X-points, respectively. 
At the O-point, the velocity distribution is close to one-peak, while a symmetric two peak distribution appears at the X-point. 
Figure \ref{figure7} shows the velocity distribution of ions with $m_\mathrm{i}=16m_\mathrm{e}$ at the same period. 
On the contrary to the electrons', a two peak distribution appears even at the O-point.
The distribution implies that ions  have not been thermalized yet due to the large mass. 
Figure \ref{figure8} shows the electron velocity distribution at the same period for the case of $m_\mathrm{i}=m_\mathrm{e}$. 
Because of the mass symmetry, the ion velocity distribution becomes identical with that of electrons by transforming $v_z$ as $v_z\rightarrow -v_z$. 
The shape of the distribution in Figure \ref{figure8} is similar to that in Figure \ref{figure6}. 
\subsection{The configuration of the magnetic field}
One can clearly see that the topology of the magnetic field changes when the filaments coalesce. 
In the framework of the resistive MHD, the coalescence of the self-confined plasma cylinders leads to the reconnection of the magnetic field \citep[][]{p57,s58,p64}, which is followed by the conversion of the magnetic energy into the kinetic energy of particles \citep[see][]{uk00,k04}. 
However, results of our simulation do not show such exchange of the energies. 
We have checked the velocity distribution around the X-point, where the magnetic reconnection might occur and found that it does not deviate from the Maxwell-Boltzmann distribution.
This is at variance with the prediction from the theory of the magnetic reconnection. 
The lack of the release of the magnetic energy seems to be attributed to the absence of the reconnection layer between the filaments, where the anti-parallel magnetic fields coexist in a narrow and compressed region, does not form between the filaments.
Instead, magnetic fields vanish at the middle point between the filaments. 

\section{QUASI-EQUILIBRIUM AND ITS TRANSITION}
From the analysis of results of the simulation, it is found that there is a quasi-equilibrium after the formation of the filamentary structure in the saturation phase of the Weibel instability. 
In this section, we consider whether the equilibrium is expressed by an analytical solution of the Vlasov-Maxwell system. 

\subsection{The analytical expression}
\begin{figure}
\begin{center}
\includegraphics[scale=0.5]{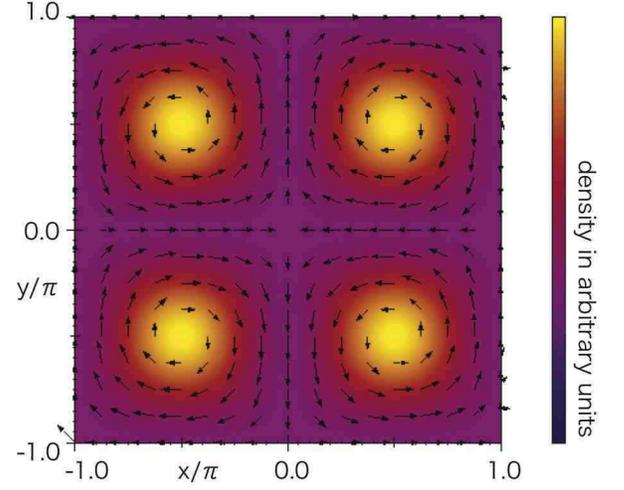}
\caption{The configuration of the stationary solution described by Equations (\ref{electron}) and (\ref{potentials}). The color codes and arrows represent the density distribution of electrons(or ions) and the direction of the magnetic field in the x-y plane, respectively.}
\label{figure9}
\end{center}
\end{figure}

\cite{ss08} present a method to construct stationary solutions of the Vlasov-Maxwell system and derive a new two-dimensional equilibrium configuration of collisionless plasmas, whose velocity distribution really resembles the quasi-equilibrium shown above. 
Slightly modifying the result of \cite{ss08}, we construct the following equilibrium, 
\begin{eqnarray}
f_e&=&C\left(v_z-A_z\right)^2\exp\left[-\frac{v_x^2+v_y^2+v_z^2}{v_e^2}\right],\label{electron}\\
f_i&=&C\left(v_z+A_z\right)^2\exp\left[-\frac{v_x^2+v_y^2+v_z^2}{v_i^2}\right],
\label{distribution}
\end{eqnarray} 
with the electromagnetic potentials expressed as
\begin{equation}
\phi=A_x=A_y=0,\ \ \ A_z=B_0\sin x\sin y,
\label{potentials}
\end{equation}
which lead to 
\begin{eqnarray}
E_x=E_y=E_z=0,
\nonumber\\
B_x=B_0 \sin x\cos y,\ \ \ B_y=-B_0\cos x\sin y,\ \ \ B_z=0.
\label{Beq}
\end{eqnarray}
One can easily check that these expressions satisfy the Vlasov-Maxwell system exactly. 

In the following we consider the case of $m_\mathrm{i}=m_\mathrm{e}$ because ions are not thermalized in the case of $m_\mathrm{i}=16m_\mathrm{e}$.  
If the stationary solution described above actually corresponds to the quasi-equilibrium, free parameters in the solution must be determined from results of the simulation. 
The value of $v_\mathrm{e}(=v_\mathrm{i})$ is derived by fitting a gaussian to the $x,y,v_x,v_z$-integrated velocity distribution of electrons in the quasi-equilibrium phase. 
Then, we obtain $v_\mathrm{e}=0.14$. 
The value of $B_0$ is derived from the magnetic energy $E_\mathrm{m}$ 
using Equations (\ref{Em}) and (\ref{Beq}) as
\begin{eqnarray}
E_\mathrm{m}&=&\frac{B_0^2}{2}\int^{\pi}_{-\pi}dx\int^{\pi}_{-\pi}dy(\sin^2x\cos^2y+\cos^2x\sin^2y)
\\&=&\pi^2B_0^2.
\end{eqnarray}
On the other hand, $E_\mathrm{m}\simeq 0.2$ from the result. 
So we obtain $B_0\simeq \sqrt{0.2}/\pi\simeq 0.14$. 
\begin{figure}
\begin{center}
\includegraphics[scale=0.5]{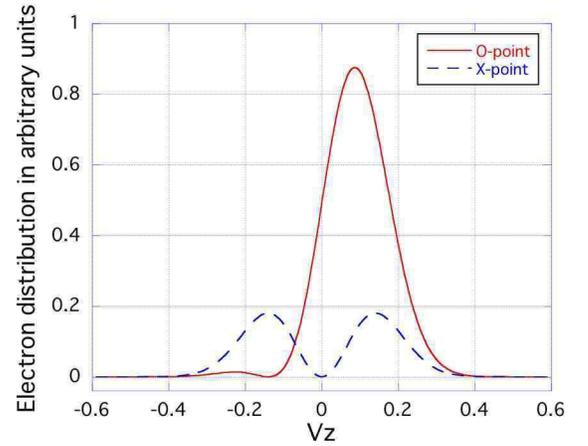}
\caption{The $v_x,v_y$-integrated electron velocity distribution of the stationary solution described by Equations (\ref{electron}) and (\ref{potentials}).  The solid and dashed lines represent the distributions at the O, and X-points, respectively.}
\label{figure10}
\end{center}
\end{figure}

Figure \ref{figure9} shows the density distribution and the configuration of the magnetic field $B_x$ and $B_y$ of the equilibrium described above.
In Figure \ref{figure10}, the solid and dashed line represent the $v_x,v_y$-integrated velocity distributions given by Equations (\ref{electron}) and (\ref{potentials}) at the O- and X-points with the parameters; $v_e=0.14$, and $B_0=0.14$. 
The filamentary structure and the configuration of the magnetic field in Figure \ref{figure9} remarkably resembles those in Figure \ref{figure4}($90<t<300$). 
In addition, comparing Figure \ref{figure10} with Figure \ref{figure8}, one can see that the dominant features appearing in Figure \ref{figure8}, the one-peak distribution at the O-point and the symmetric double-peak distribution at the X-point, are well reproduced by the analytical expressions (\ref{electron}) and (\ref{potentials}). 

\begin{figure}
\begin{center}
\includegraphics[scale=0.5]{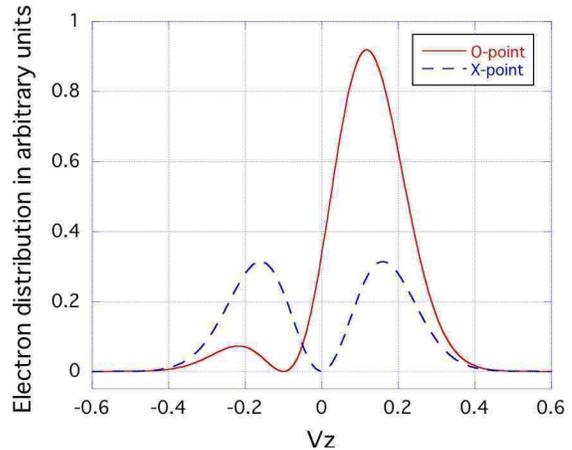}
\caption{Same as Figure \ref{figure10}, but for $v_e=0.16$ amd $B_0=0.1$.}
\label{figure11}
\end{center}
\end{figure}

In the above discussion, we use the parameter $v_e$ and $B_0$ determined from the result of our simulation. 
However, the shape of the velocity distribution is sensitive to the parameters. 
The velocity distribution with different parameters, $v_e=0.16$ and $B_0=0.1$, is shown in Figure \ref{figure11}, which is more similar to the result of our simulation shown in Figure \ref{figure8}. 

Therefore, we conclude that the equilibrium configuration given by Equations (\ref{electron}) and (\ref{potentials}) provides an appropriate expression of the quasi-equilibrium achieved in the beginning of the saturation phase, as long as the velocity distribution of electrons or that of ions with $m_\mathrm{i}=m_\mathrm{e}$ are concerned.

\subsection{Remaining problems}
Although we see that there is an analytical expression to describe the quasi-equilibrium configuration, some problems remain.  

First problem is the disagreement of the distribution of ions with $m_\mathrm{i}=16m_\mathrm{e}$ between the analytical expression and the result of the simulation. 
As in Figure \ref{figure7}, the velocity distribution of ions with $m_\mathrm{i}=16m_\mathrm{e}$ have sharp peaks compared with the electron velocity distribution, which cannot be reproduced by the expressions (\ref{distribution}) and (\ref{potentials}). 
This disagreement arises since we truncate the series of $v_z^n\exp(-v_z^2/v_j^2),\ (n=0,1,2,\cdots)$ at the second-order ($n=2$) in constructing the velocity distribution (\ref{distribution}). 
So this problem can be resolved by constructing higher-order stationary solutions of the Vlasov-Maxwell system according to the procedure given by \cite{ss08};
\begin{eqnarray}
f_{j}&\propto& \left[g_{j,0}+g_{j,1}H_1(v_z/v_j)+g_{j,2}H_2(v_z/v_j)\right.
\nonumber\\&&
\left.+g_{j,3}H_3(v_z/v_j)+\cdots\right]
\exp\left[-\frac{v_x^2+v_y^2+v_z^2}{v_j^2}\right],
\end{eqnarray}
where $H_n(v_z/v_j)$ are Hermite polynomials. Their coefficients $g_{j,n}$ are functions of $A_z$. 
While this prescription enables us to make a more rigorous fits to the results of the simulation, the magnetic field cannot be expressed analytically. 
Furthermore, the expressions (\ref{distribution}) and (\ref{potentials}) are sufficient to reproduce the main feature of the result of the simulation, i.e., a single peak velocity distribution at the O-point and a symmetric double peak velocity distribution at the X-point. 
Therefore we take the construction of the higher-order stationary solutions as beyond the scope of the present work.

Second is on the stability of the equilibrium. 
Results of the numerical simulation with the scale length of $2\pi$ reveal that the quasi-equilibrium configuration achieved in the beginning of the saturation phase is unstable. 
The transition from the quasi-equilibrium to another appears in Figure \ref{figure4}. 
This clearly indicates that the equilibrium configuration expressed by Equations (\ref{electron})-(\ref{Beq}) is unstable. 
However, there is no analytical or numerical investigation into its stability and the long-term evolution. 
It is not clear from our simulation whether the critical scale length above which the equilibrium is stable exists or not. 
Nevertheless, we could make a brief speculation on the transition from the quasi-equilibrium state based on the result of the simulation. 
Comparing the time evolution of the kinetic energies $K_{ex}$ and $K_{ez}$ shown in Figure \ref{figure3}, and the time evolution of the filamentary structure shown in Figure \ref{figure4}, we can see a coincidence. 
At the beginning of the saturation phase ($200<t<300$), $K_{ez}$ takes greater values than $K_{ex}$. 
Then, after the quasi-equilibrium configuration breaks, $K_{ex}$ and $K_{ez}$ oscillate changing their magnitude relation. 
Eventually, the amplitude of the oscillation decays and then $K_{ex}$ and $K_{ez}$ take similar values.
This behavior coincides with the oscillation and the coalescence of the two filaments. 
In other words, the coalescence of filaments and the dissolution of the anisotropic distribution of the kinetic energies, $K_{ex}$ and $K_{ez}$, occur simultaneously.  

\section{DISCUSSIONS AND CONCLUSIONS}
In this paper, we have presented the behavior of a filamentary structure resulting from the Weibel instability. 
In order to simulate the behavior with high accuracy, we use a numerical scheme for the direct integration of the Vlasov-Maxwell system. 
The results reveal that there is a quasi-equilibrium configuration after the saturation of the Weibel instability.
Then, it turns into another configuration. 
The quasi-equilibrium configuration is found to be described by the analytical expression (\ref{electron})-(\ref{Beq}). 
The coalescence of the filaments is synchronized with the time evolution of the kinetic energies of electrons $K_{ex}$ and $K_{ez}$. 

Comparing results of the simulation, we consider the validity of models proposed so far. 
\cite{m06} treated the filamentary structure in the framework of the MHD. 
In our simulation, however, the velocity distribution at the X-point is far from the Maxwell-Boltzmann distribution as seen in Figures \ref{figure6}, \ref{figure7}, and \ref{figure8}, which implies the treatment of the plasma as a magnetized fluid is not appropriate. 
Furthermore, although they assumed that the filament is isolated and other filaments assert a negligible effect,  two peaks in the velocity distribution at the X-point resultant from our simulation show that the outskirts of filaments coexist around the X-point. 
The model proposed by \cite{m05} is based on a toy model. 
Although this model ignores the effect of the dissipation of the magnetic field energy via reconnection of the magnetic fields, it is expected in the framework of the resistive MHD. 
Our simulation does not show any sign of the reconnection. 
As stated earlier, we attribute the lack of the magnetic reconnection to the absence of the reconnection layer between the filaments. 
But, this can also be due to the coarseness of the simulation. 
While judging whether the lack of the magnetic reconnection is real or not needs careful investigation, the results of our simulation justify the treatment of the coalescence in \cite {m05}. 
\cite{csa08} modeled the decay of the magnetic turbulence generated by the Weibel instability. 
This treatment is appropriate only in the region where the magnetic trapping is not so important. 
Actually, \cite{csa08} confirm that their model cannot predict the time evolution of  magnetic field strength in the presence of a magnetic field with long wavelength. 

While the detailed dynamical behavior of the filamentary structure is revealed in this study, some problems, for example, the long-term evolution of the filaments and whether the critical scale-length above which the equilibrium is stable exists or not, remain unclear. 
In addition, because the distribution function is assumed to be uniform along the direction of the initial bulk flow of plasmas, we cannot deal with some 3D perturbation on the equilibrium, for example, bending and twisting of the filaments. 
The stability of the equilibrium described by Equations (\ref{electron})-(\ref{Beq}), which is presumably a key ingredient for a strong magnetic field to survive in long term and serve as an accelerator of particles in the transition layer of a collisionless shock front, should be investigated rigorously including 3D effects. 

Although this study treats non-relativistic plasmas, relativistic effects become vital in some astrophysical phenomena, e.g., GRB afterglows, jets from AGN. 
\cite{s08} provides the relativistic extension of \cite{ss08}, i.e, stationary solutions of the relativistic Vlasov-Maxwell system. 
It is not irrelevant that one expects the equilibrium constructed by \cite{s08} to provide an appropriate model for the filamentary structure resulting from the Weibel instability in relativistic plasmas as the non-relativistic counterpart does. 

\end{document}